\documentclass{article}    
\usepackage{makeidx}
\usepackage{amssymb}
\usepackage{amsfonts}
\usepackage{amsmath}
\usepackage{amsthm}

\newtheorem{theorem}{Theorem}[section]
\newtheorem{lemma}[theorem]{Lemma}
\newtheorem{proposition}[theorem]{Proposition}

\newtheorem{definition}[theorem]{Definition}

\begin{document}

\title{A Phase Transition in a Quantum
Crystal with Asymmetric Potentials\thanks{Supported by the DFG
through the Project 436 POL 113/115/0-1}} \vskip1cm
\author{Alina Kargol and Yuri Kozitsky \\[.5cm]
{\sf akargol@golem.umcs.lublin.pl} \&  {\sf
jkozi@golem.umcs.lublin.pl}\\[.5cm] Instytut Matematyki,\\[.3cm] Uniwersytet
Marii Curie-Sk{\l}odowskiej, \\[.3cm] Lublin 20-031 Poland  }

\maketitle \noindent

\begin{abstract}
A translation invariant system of interacting quantum anharmonic
oscillators indexed by the elements of a simple cubic lattice
$\mathbb{Z}^d$ is considered. The anharmonic potential is of general
type, which in particular means that it might have no symmetry. For
this system, we prove that the global polarization (obtained in the
thermodynamic limit) gets discontinuous at a certain value of the
external field  provided $d\geq 3$, and the particle mass as well as
the interaction intensity are big enough. The proof is based on the
representation of local Gibbs states in terms of path measures and
thereby on the use of the infrared estimates and the
Garsia-Rodemich-Rumsey inequality.
\end{abstract}


\section{Introduction and Setup}

The theory of phase transitions in quantum systems  has essential
peculiarities, which distinguish it from the corresponding theory of
classical systems. In this context, it suffices to mention that the
existence of phase transitions in the three-dimensional isotropic
quantum Heisenberg model has not been proven yet. For lattice
models, most of the results in this domain were obtained by means of
quantum versions of the method of infrared bounds developed in
\cite{FSS}. The first publication in which the infrared estimates
were applied to quantum spin models seems to be the article
\cite{DLS}. After certain modifications this method was applied to a
number of models with unbounded Hamiltonians \cite{DLP,PK,BK,Kond},
the main characteristic feature of which was the $Z_2$-symmetry
broken by the phase transition. This symmetry allowed for obtaining
an estimate crucial for the method. However in classical models, for
proving phase transitions by means of the infrared estimates,
symmetry was not especially important, see Theorem 3.5 in \cite{FSS}
and the discussion preceding this theorem. There might be two
explanations of such a discrepancy: (a) the symmetry was the key
element, but only of the methods employed therein, and, like in the
classical case, its lack does not imply the lack of phase
transitions; (b) the symmetry is crucial in view of e.g. quantum
effects, which stabilize the system, see \cite{Alb4,Alb5}. So far,
there has been no possibility to check which of these explanations
is true. In this letter, we prove that the system of interacting
quantum anharmonic oscillators without any symmetry undergoes a
phase transition if in particular the oscillator mass is
sufficiently big and hence  quantum effects are not so strong. As a
consequence, the dilemma mentioned above has been solved in favor of
explanation (a).

An infinite system of interacting quantum anharmonic oscillators
indexed by the elements of a crystal lattice is called (with a
certain abuse of terminology) {\em a quantum anharmonic crystal}.
Mostly such models are related with ionic crystals containing
localized light particles oscillating in the field created by heavy
ionic complexes, see \cite{St}. An example here can be a ${\rm
KDP}$-type ferroelectric with hydrogen bounds, in which the
particles are protons or deuterons performing one-dimensional
oscillations along the bounds.  To models of this kind the method of
infrared estimates was first applied in \cite{DLP,PK}, where the
anharmonic potential was $Z_2$-symmetric and of $\phi^4$ type. These
two properties allowed for obtaining the crucial estimate by means
of a purely quantum tool - the Bogoliubov inequality. Afterwards in
\cite{BK,Kond}, the method of infrared estimates was extended to
cover the case of $Z_2$-symmetric anharmonic potentials of general
type, which have two deep enough wells. This was achieved my means
of a representation of the Gibbs states in the form of
imaginary-time Feynman path integrals. The approach in quantum
statistical physics based on such a representation is called {\em
Euclidean} due to its conceptual similarity with the corresponding
approach in quantum field theory, see \cite{[Sim1]}.
 In this
approach, the model is treated as a system of interacting classical
spins, which are infinite-dimensional and unbounded. Thereby, the
method of \cite{FSS} can be applied directly if the estimate
mentioned above is obtained.

In this letter, we prove the existence of phase transitions in
quantum anharmonic crystals with possibly asymmetric anharmonic
potentials. We employ an updated version \cite{Alb3,Alb6,KP} of the
Euclidean method used in \cite{BK,Kond}, combined with some new
techniques. Namely, we prove an analog of Lemma 3.4 of \cite{FSS} by
means of the Garsia-Rodemich-Rumsey lemma, which then is used to
prove an analog of Theorem 3.5 of \cite{FSS}.

The heuristic Hamiltonian of the model we consider is
\begin{equation} \label{1}
H = - \frac{J}{2} \sum_{\ell,\ell': \ |\ell - \ell'|=1}  q_{\ell}
q_{\ell'} + \sum_{\ell} H_{\ell},
\end{equation}
where the sums run through a lattice ${\mathbb{ L}}={\mathbb{Z}}^d$
and the displacement $q_\ell$ is one-dimensional. The interaction
term is of dipole-dipole type; we assume that $J>0$. The Hamiltonian
\begin{equation} \label{2} H_\ell = H_\ell^{\rm har} + V (q_\ell
) \ \stackrel{\rm def}{=} \ \frac{1}{2m} p_\ell^2 + \frac{a}{2}
q_\ell^2 + V (q_\ell ), \quad a>0,
\end{equation}
 describes an isolated anharmonic oscillator of mass
$m$ and momentum $p_\ell$, whereas $H^{\rm har}_\ell$ corresponds to
a quantum harmonic oscillator of rigidity $a$. Regarding the
anharmonic potential, we assume that it contains an external field
$h\in \mathbb{R}$ and is of the form
\begin{equation}
\label{5A} V(x) = V_0(x) - h x,
\end{equation}
where $V_0$ is continuous, such that $V_0 (0) = 0$, and for all
$x\in \mathbb{R}$,
\begin{equation} \label{5}
A_V x^{2r} + B_V \leq V_0 (x) ,
\end{equation}
with certain $r>1$, $A_V > 0$, $B_V\in \mathbb{R}$. Like all objects
of this kind, the Hamiltonian (\ref{1}) is `represented' by local
Hamiltonians corresponding to non-void finite subsets
$\Lambda\subset \mathbb{L}$. For such $\Lambda$, we write $\Lambda
\Subset \mathbb{L}$; by $|\Lambda|$ we denote the number of
elements. The adjective {\em local} will always mean the property of
being related with a certain $\Lambda \Subset \mathbb{L}$, whereas
{\em global} will refer to the whole lattice. By $(\cdot , \cdot)$
and $|\cdot|$ we denote the scalar product and norm in
$\mathbb{R}^d$.

The set $\{ \Lambda \}_{ \Lambda \Subset \mathbb{L}}$ is countable;
it is a net with the order defined by inclusion. A linearly ordered
sequence of subsets $\Lambda \Subset \mathbb{L}$, which exhausts the
lattice $\mathbb{L}$, will be called a cofinal sequence.  The limit
of a sequence of appropriate $A_\Lambda$ taken along a cofinal
sequence $\mathcal{L}$ will be denoted by $\lim_{\mathcal{L}}
A_\Lambda$; we write $\lim_{\Lambda \nearrow \mathbb{L}} A_\Lambda$
if the limit is taken along $\{ \Lambda \}_{ \Lambda \Subset
\mathbb{L}}$. The same notations will be used for $\lim\sup$ and
$\lim\inf$.

 Thereby, the local Hamiltonian is
\begin{equation}
\label{LH} H_\Lambda = - \frac{J}{2}\sum_{\ell, \ell'\in \Lambda: \
|\ell - \ell'|=1} q_\ell q_{\ell'} + \sum_{\ell\in
\Lambda}\left[H_{\ell}^{\rm har} + V(q_\ell) \right].
\end{equation}
A special kind of $\Lambda \Subset \mathbb{L}$ is the box
\begin{equation} \label{8}
\Lambda = (- L , L]^d \bigcap \mathbb{L}, \quad L \in \mathbb{N},
\end{equation}
which can be turned into a torus by setting periodic conditions on
its boundary. The same can be done by equipping $\Lambda$ with the
periodic distance $|\ell - \ell'|_\Lambda$, the definition of which
is standard. By $\mathcal{L}_{\rm box}$ we denote the set of all
boxes. For a box $\Lambda$, we set
\begin{equation} \label{9}
H^{\rm per}_\Lambda = - \frac{J}{2}\sum_{\ell, \ell'\in \Lambda: \
|\ell - \ell'|_\Lambda=1} q_\ell q_{\ell'} + \sum_{\ell\in
\Lambda}\left[H_{\ell}^{\rm har} + V(q_\ell) \right],
\end{equation}
that is the periodic local Hamiltonian invariant with respect to the
translations of the torus $\Lambda$. In the sequel, by writing
expressions like $H^{\rm per}_\Lambda $ we tacitely assume that
$\Lambda$ is a box.  Due to (\ref{5}), both $H_\Lambda$ and $H^{\rm
per}_\Lambda$ are self-adjoint operators in the Hilbert space
$\mathcal{H}_\Lambda = L^2 (\mathbb{R}^{|\Lambda|})$, such that
 for every $\beta>0$,
\begin{eqnarray} \label{10}
Z_\Lambda  & \stackrel{\rm def}{=} & {\rm trace}\left[ \exp (- \beta
{H}_\Lambda)\right] < \infty, \\  \quad Z^{\rm per}_\Lambda  &
\stackrel{\rm def}{=} & {\rm trace}\left[ \exp (- \beta {H}^{\rm
per}_\Lambda)\right] < \infty. \nonumber
\end{eqnarray}
Since the inverse temperature plays no role in our constructions, we
set $\beta =1$. We also set $\hbar = 1$.
 Thereby, we define the local Gibbs states
\begin{eqnarray} \label{11}
\varrho_{\Lambda} (A) & = & {\rm trace}\left[ A\exp (-
{H}_\Lambda)\right]/ Z_\Lambda, \quad A\in \mathfrak{C}_\Lambda, \\
\varrho^{\rm per}_{ \Lambda} (A) & = & {\rm trace}\left[ A\exp (-
{H}^{\rm per}_\Lambda)\right]/ Z^{\rm per}_\Lambda. \nonumber
\end{eqnarray}
Here $\mathfrak{C}_\Lambda$ is the local algebra of observables,
consisting  of all bounded linear operators on
$\mathcal{H}_\Lambda$. We study the dependence of the averages
(\ref{11}) on $J$, and $h$. Among them are local
\emph{polarizations}
\begin{equation} \label{12}
M_\Lambda^{\rm per} (J,h) = \varrho^{\rm per}_{ \Lambda} (q_\ell),
\qquad M_{\Lambda} (J,h) = \frac{1}{|\Lambda|} \sum_{\ell \in
\Lambda}\varrho_{ \Lambda} (q_\ell).
\end{equation}
\begin{proposition} \label{1pn}
Both $\{M_\Lambda (j,h)\}_{\Lambda\Subset \mathbb{L}}$ and $\{M^{\rm
per}_\Lambda (j,h)\}_{\Lambda\in \mathcal{L}_{\rm box}}$
 are bounded.
\end{proposition}
Set
\begin{eqnarray}
\label{12C} M_{+} (J,h) & = & \max\left\{\lim\sup_{\Lambda \nearrow
\mathbb{L}} M_\Lambda (J,h) \ ;
\ \lim\sup_{\Lambda \in \mathcal{L}_{\rm box}} M^{\rm per}_\Lambda (J,h)\right\},\qquad \\
M_{-} (J,h)& = & \min\left\{\lim\inf_{\Lambda \nearrow \mathbb{L}}
M_\Lambda (J,h) \ ; \   \lim\inf_{\Lambda \in \mathcal{L}_{\rm box}}
 M^{\rm per}_\Lambda (J,h)\right\}.
\nonumber
\end{eqnarray}
In the following, by a denumerable set we mean the set which is
void, finite, or countable.
\begin{proposition} \label{2pn}
For every fixed $J$, there exists a denumerable set
$\mathcal{R}\subset \mathbb{R}$, such that, for $h \in \mathcal{R}^c
\ \stackrel{\rm def}{=} \ \mathbb{R}\setminus \mathcal{R}$,
\begin{equation}
\label{12D}
 M_{-} (J,h) = M_{+} (J,h) \ \stackrel{\rm def}{=} \   M(J,h).
\end{equation}
The polarization $M(J,h)$, as a function of $h$, is nondecreasing on
$\mathcal{R}^c$; it is continuous on each its open connected
component.
\end{proposition}
Note that by $\mathcal{R}$ we mean the smallest set with the
properties stated.

By Proposition \ref{1pn}, it follows that for a specific cofinal
sequence $\mathcal{L}$, which may also be composed by boxes, the
corresponding sequence of local polarizations has a limit in $[
M_{-} (J,h), M_{+} (J,h)]$. By Proposition \ref{2pn}, this interval
shrinks into a point if $h\in \mathcal{R}^c$, which merely means
that, at such $h$, there exists a (global) polarization, independent
of the sequence $\mathcal{L}$ along which the thermodynamic limit
has been taken. This polarization is continuous on the interval
$(a_{-}, a_{+})\subset \mathcal{R}^c$, where $a_{\pm}$ are two
consecutive elements of $\mathcal{R}$. At such $a_{\pm}$, $M(J,h)$
is discontinuous.  Therefore, at each $a\in \mathcal{R}$, such that
both $(a-\epsilon, a)$, $(a, a+\epsilon)$ are subsets of
$\mathcal{R}^c$ for a certain $\epsilon >0$,  one has
\begin{equation} \label{12E}
\lim_{h\uparrow a} M(J,h) <  \lim_{h\downarrow a} M(J,h) .
\end{equation}
At the same time, the set $\mathcal{R}^c$ may have empty interior;
hence, the global polarization may be nowhere continuous.
\begin{definition} \label{1df}
 The model considered undergoes a phase transition (of
first order) at certain  $J$ and $h_*$ if  the global polarization,
as a function of $h$ at fixed $J$, is discontinuous at this $h_*$.
\end{definition}
Note that this definition agrees with the known one given by L.
Landau and we do not suppose that the phase transition breaks any
symmetry.
\begin{theorem} \label{1tm}
For every $d\geq 3$, there exist $m_*>0$ and $J_*>0$ such that for
every $m>m_*$ and $J>J_*$, there exists $h_*\in \mathbb{R}$,
possibly dependent on $m$ and $J$, such that $M( J,h)$ gets
discontinuous at $h_*$, i.e., the model undergoes a phase
transition.
\end{theorem}

\section{Proof of the Theorem}
\label{PT}

The proof of Theorem \ref{1tm} is based on a number of preparatory
statements -- Propositions and Lemmas. Propositions will be either
taken from other sources or discussed in the subsequent parts of the
article. Some of them will be proven here. Lemmas are proven in
Section \ref{sec4}.

Define
\begin{equation}
\label{12F} p_\Lambda (J,h) = \log Z_\Lambda/|\Lambda|; \qquad
p^{\rm per}_\Lambda (J,h) = \log Z^{\rm per}_\Lambda/|\Lambda|.
\end{equation}
We shall call these functions {\em the  local pressure} and {\em the
periodic local pressure} respectively. Their properties are
described by the next two statements. The first one follows directly
from the definition (\ref{12F}).
\begin{proposition}
\label{3pn} Each $p_\Lambda (J,h)$, $p^{\rm per}_\Lambda (J,h)$ is
an infinitely differentiable function of $h\in \mathbb{R}$ and $J\in
\mathbb{R}$. It is also a convex function of both variables.
\end{proposition}
One can easily verify that
\begin{equation}
\label{12G} \frac{\partial}{\partial h}p_\Lambda (J,h) = M_\Lambda
(J,h), \qquad  \frac{\partial}{\partial h}p^{\rm per}_\Lambda (J,h)
= M^{\rm per}_\Lambda (J,h).
\end{equation}
The second statement clarifies the situation with the limiting
pressure.
\begin{proposition} \label{4pn}
For every $J\geq 0$ and $h\in \mathbb{R}$,
\begin{equation}
\label{12L} \lim_{\Lambda \nearrow \mathbb{L}} p_\Lambda (J,h) =
\lim_{ \mathcal{L}_{\rm box}} p^{\rm per}_\Lambda (J,h) \
\stackrel{\rm def}{=} \ p(J,h).
\end{equation}
\end{proposition}
 This fact together with those established by Proposition
\ref{3pn} yield important information about the global
polarizations. It is known, see e.g. \cite{[Simon]}, pages 34 - 37,
that for a convex function $\varphi: \mathbb{R} \rightarrow
\mathbb{R}$, \vskip.1cm
\begin{tabular}{ll}
(a)  &the one-sided derivatives $ \varphi'_{\pm}(t)$ exist for every
$t\in \mathbb{R}$;\\ &the set $\{t\in \mathbb{R} \ | \
\varphi'_{+} (t) \neq  \varphi'_{-} (t)\}$ is denumerable;\\[.2cm]
(b)  &the point-wise limit $\varphi$ of a sequence of convex
functions\\ &$\{\varphi_n\}_{n\in \mathbb{N}}$ is a convex function;
if $\varphi$ and all $\varphi_n$'s are\\ &differentiable at a given
$t$, $\{\varphi'_{n}(t)\}_{n\in \mathbb{N}}$ converges to
$\varphi'(t)$.
\end{tabular}
\vskip.1cm \noindent Thereby, the proof of Proposition \ref{2pn}
readily follows from the convexity of the pressure and the latter
facts.

In the proof of Theorem \ref{1tm}, we employ the Duhamel two-point
function - a traditional element of the theory of phase transitions
in quantum systems, see \cite{DLS}. In our context, it is
\begin{equation} \label{a}
D^\Lambda_{\ell \ell'} = \int_0^1 \int_0^1
\mathit{\Gamma}^\Lambda_{\ell \ell'} (\tau, \tau'){\rm d}\tau {\rm
d}\tau' = \int_0^1 \mathit{\Gamma}^\Lambda_{\ell \ell'} (0 ,
\tau){\rm d}\tau, \quad \ell , \ell' \in \Lambda ,
\end{equation}
where
\begin{equation}\label{12J}
  \mathit{\Gamma}^\Lambda_{\ell\ell'}(\tau,\tau')
   = \varrho^{\rm per}_\Lambda \left\{q_\ell \exp\left[- (\tau' -
\tau)H^{\rm per}_\Lambda \right] q_{\ell'} \exp\left[(\tau' -
\tau)H^{\rm per}_\Lambda \right]\right\} .   \nonumber
\end{equation}
One can show that
\begin{equation}
\label{a0} D^\Lambda_{\ell \ell'} \geq 0,
\end{equation}
for all boxes $\Lambda$ and $\ell, \ell'\in \Lambda$.
 By construction,
 $D^\Lambda_{\ell \ell'}$ is invariant
with respect to the translations of the torus $\Lambda$. Thus, we
introduce the set $\Lambda_*$ (a Brillouin zone) consisting of
\begin{equation}
\label{aa}
 p = (p_1 , \dots , p_d), \quad p_j = -
\pi + \frac{2\pi}{L} s_j, \ s_j = 1 , \dots , 2L, \ j = 1, \dots ,
d,
\end{equation}
and thereby the Fourier transformation
\begin{eqnarray}
\label{ab} \widehat{D}^\Lambda_p & = & \sum_{\ell'\in \Lambda}
D^\Lambda_{\ell \ell'} \exp[\imath (p, \ell - \ell')],\quad p \in \Lambda_* ,\\
D^\Lambda_{\ell \ell'} & = & \frac{1}{|\Lambda|} \sum_{p \in
\Lambda_*} \widehat{D}^\Lambda_p
 \exp[- \imath (p, \ell - \ell')]. \nonumber
\end{eqnarray}
Now let us make a more formal definition of the thermodynamic limit.
For $\Lambda \subset \Lambda'\Subset \mathbb{L}$, one can define the
canonical embedding $\mathfrak{C}_\Lambda \hookrightarrow
\mathfrak{C}_{\Lambda'}$, up to which $\mathfrak{C}_\Lambda$ be a
subalgebra of $\mathfrak{C}_{\Lambda'}$. Thereby, we define \[
\mathfrak{C}^{\rm loc} = \bigcup_{\Lambda \Subset \mathbb{L}}
\mathfrak{C}_\Lambda.
\]
Equipped with the norm inherited from $\mathfrak{C}_\Lambda$, this
algebra becomes a normed space. Its completion, $\mathfrak{C}$, is
called the algebra of quasi-local observables.
\begin{proposition} \label{perpn}
For every $J>0$ and $h\in \mathbb{R}$, there exists a tending to
infinity sequence $\{L_n\}_{n\in \mathbb{N}}$ and hence the sequence
$\{\Lambda_n\}_{n\in \mathbb{N}}$ of the corresponding boxes
(\ref{8}), such that the sequence $\{\varrho^{\rm
per}_{\Lambda_n}\}_{n\in \mathbb{N}}$ converges to a translation
invariant state $\varrho^{\rm per}$ on $\mathfrak{C}$ (called a
periodic state) in such a way that, for every polynomial
$P(q_\Lambda)$, $\Lambda \Subset \mathbb{L}$, $q_\Lambda =
(q_\ell)_{\ell \in \Lambda}$,
\begin{equation} \label{abba0}
\lim_{n\rightarrow +\infty}\varrho^{\rm per}_{\Lambda_n}
[P(q_\Lambda)] = \varrho^{\rm per} [P(q_\Lambda)].
\end{equation}
Furthermore, for every $\ell, \ell'$, $\mathit{\Gamma}_{\ell
\ell'}^{\Lambda_n} \rightarrow \mathit{\Gamma}_{\ell \ell'}$,
 uniformly on $\tau, \tau' \in [0,1]$.
\end{proposition}
Note that by (\ref{abba0}), one has, c.f. (\ref{12C}),
\begin{equation} \label{abk}
\varrho^{\rm per} (q_\ell) = \lim_{n\rightarrow +
\infty}M_{\Lambda_n}^{\rm per}(J,h).
\end{equation}
Set, c.f. (\ref{a}),
\begin{equation}
\label{abba1} D_{\ell \ell'} = \int_0^1 \mathit{\Gamma}_{\ell \ell'}
(0,\tau){\rm d}\tau.
\end{equation}
By construction, $\mathit{\Gamma}_{\ell \ell'}$, and hence $D_{\ell
\ell'}$, are invariant with respect to the translations of
$\mathbb{L}=\mathbb{Z}^d$. Thus, one can define
\begin{eqnarray}
\label{ac} \widehat{D}_p & = & \sum_{\ell'} D_{\ell \ell'}
\exp[\imath (p, \ell - \ell')],\quad p \in (-\pi, \pi]^d, \\
D_{\ell \ell'} & = & \frac{1}{(2\pi)^d} \int_{(\pi, \pi]^d}
\widehat{D}_p \exp[-\imath (p, \ell - \ell')]{\rm d}p. \nonumber
\end{eqnarray}
\begin{proposition} \label{abbalm}
Suppose that for a given $J>0$, there exists $h_*$ such that
\begin{equation} \label{b}
\varrho^{\rm per} (q_\ell ) = 0
\end{equation}
at $h=h_*$. Suppose in addition that at $h=h_*$ there exists a
sequence of boxes $\{\Lambda_n\}_{n\in \mathbb{N}}$ such that
\begin{equation}
\label{ba} \lim_{n\rightarrow + \infty}
 \frac{1}{|\Lambda_n |} \sum_{\ell'\in \Lambda_n} D_{\ell \ell'} =
\lim_{n\rightarrow + \infty} \frac{1}{|\Lambda_n|^2}
\sum_{\ell,\ell'\in \Lambda_n} D_{\ell \ell'} >0.
\end{equation}
Then the model undergoes the phase transition at these $J$ and
$h_*$.
\end{proposition}
By (\ref{ac}) and (\ref{a0}) it follows from (\ref{ba}) that
$\widehat{D}_p$ is singular at $p=0$ in this case.
 On the other hand, by the second line
of (\ref{ac}), $\widehat{D}_p$ is a distribution; hence, one can
write
\begin{equation}\label{ba2}
\widehat{D}_p = (2 \pi)^d \varkappa \delta (p) + g(p),
\end{equation}
where $\delta$ is the Dirac $\delta$-function and $g(p)$ is regular
at $p=0$.
 By (\ref{a0}), $g(p)$, for all $p$, and
$\varkappa$ are nonnegative; $\varkappa$ is positive if (\ref{ba})
holds. By (\ref{ac}) and (\ref{ba2}),
\begin{equation} \label{ba3}
\varkappa = D_{\ell \ell} - \frac{1}{(2\pi)^d} \int_{(-\pi, \pi]^d}
g(p) {\rm d}p.
\end{equation}
Thereby, in order to prove that $\varkappa >0$ one has to estimate
$D_{\ell \ell}$ from below and $g(p)$ from above. The latter
estimate  is obtained in the next statement which can be proven by
means of a method used in \cite{DLS}, see Example 4, pages 362 -
364.
\begin{proposition} \label{apn}
Suppose there exists a continuous function $b:  (-\pi,
\pi]^d\setminus \{0\} \rightarrow [0,+\infty)$ satisfying the
condition
\begin{equation}
\label{ad} \int_{(-\pi, \pi]^d} b(p) {\rm d}p < \infty ,
\end{equation}
and such that for all boxes $\Lambda$,
\begin{equation}
\label{ad1} \widehat{D}^\Lambda_p \leq b(p) , \quad \ \
 {\rm for} \ \ {\rm all} \ \ p \in \Lambda_* \setminus \{0\}.
\end{equation}
Then the function $g$  obeys the estimate
\begin{equation}
\label{ad2} g(p) \leq b(p) , \quad \ \ {\rm for} \ \ {\rm all} \ \
p\in (-\pi, \pi]^d\setminus \{0\}.
\end{equation}
\end{proposition}
A concrete form of the function $b$ is obtained by the infrared
estimates. A detailed presentation of the corresponding method in
its application to quantum anharmonic crystals is given in
\cite{BK,Kond,Alb3}, where one can find the following
\begin{proposition} \label{ibpn}
For every box $\Lambda$, and any $p\in \Lambda_* \setminus \{0\}$,
\begin{equation}\label{c}
0 < \widehat{D}^\Lambda_p \leq  1/J E(p),
\end{equation}
  where
\begin{equation}
\label{c1} E(p) = \sum_{j=1}^d [1 - \cos p_j].
\end{equation}
\end{proposition}
Note that the function $1/E(p)$ is integrable on $(-\pi, \pi]^d$ for
$d\geq 3$.

Now we give the statements which finalize the preparation of the
proof of the theorem.
\begin{lemma} \label{c3lm}
For every $m_0>0$, there exist $h_{\pm}(m_0)\in \mathbb{R}$, $h_+
(m_0)
> h_{-} (m_0)$, such that for all $m>m_0$ and $J\geq 0$,
\begin{eqnarray}
\label{c4} M^{\rm per}_\Lambda (J, h) & > & 0, \quad {\rm for} \ \
{\rm all} \ \ h>h_{+}(m_0); \\ M^{\rm per}_\Lambda (J, h) & < & 0,
\quad {\rm for} \ \ {\rm all} \ \ h<h_{-}(m_0). \nonumber
\end{eqnarray}
\end{lemma}
The next statement is an analog of Lemma 3.4 of \cite{FSS}.
\begin{lemma}
\label{c1lm}
 There exist positive $\varepsilon$, $\delta$,
and $m_*$, such that for all $\Lambda$ and $m>m_*$,
\begin{equation}
\label{c2} p^{\rm per}_\Lambda (J,h) -p^{\rm per}_\Lambda (0,h) \geq
d (\varepsilon J - \delta).
\end{equation}
\end{lemma}
\begin{lemma} \label{c2lm}
Let $m_*$ be as above. Then for every $\Lambda$, $m>m_*$, $J>0$, and
$h \in \mathbb{R}$,
\begin{equation}
\label{c3} \varrho^{\rm per}_\Lambda (q_\ell^2) \geq [p^{\rm
per}_\Lambda (J,h) -p^{\rm per}_\Lambda (0,h)]/ Jd.
\end{equation}
 \end{lemma}
One observes that $p^{\rm per}_\Lambda (0,h)$ does not depend on
$\Lambda$. By means of Lemmas \ref{c1lm}, \ref{c2lm} we have
\begin{equation}
\label{c5} \varrho_\Lambda^{\rm per} (q_\ell^2) \geq \varepsilon -
\delta/J,
\end{equation}
which can be used to estimate $D_{\ell\ell}$ from below. Note that
the lower bound of $\varrho_\Lambda^{\rm per} (q_\ell^2)$ was
mentioned in the introduction as a crucial estimate; in
\cite{DLP,PK} it was derived by the Bogoliubov inequality.

In our version of the infrared bound method, we estimate the Duhamel
function, which is performed   by means of (\ref{c5}) and
 the Bruch-Falk inequality, see Theorem 3.1 in \cite{DLS} or Theorem IV.7.5, page
392 of \cite{[Simon]}.
\begin{proposition} [Bruch-Falk Inequality] \label{bfpn}
It follows that
\begin{equation}
\label{c6} D^\Lambda_{\ell \ell} \geq \varrho_\Lambda^{\rm per}
(q_\ell^2)
 \cdot f\left(\frac{1}{4 m  \varrho_\Lambda^{\rm per} (q_\ell^2)} \right),
\end{equation}
where $f:[0,+\infty) \rightarrow [0,1]$ is defined implicitly by
\begin{equation}
\label{c7} f ( \xi \tanh \xi) = \xi^{-1} \tanh \xi, \quad {\rm for}
\ \ \xi >0; \quad {\rm and} \  \ f(0) =1.
\end{equation}
\end{proposition}
 As the right-hand side of (\ref{c6})
is independent of $\Lambda$, one can pass here to the thermodynamic
limit along the same sequence of boxes as in (\ref{abk}). Then
\begin{equation}
\label{c8a} D_{\ell \ell} \geq (\varepsilon - \delta/J)
 \theta (J); \qquad \theta (J) \ \stackrel{\rm def}{=} \
f (J/ 4 m_* (\varepsilon J - \delta)),
\end{equation}
where $m_*$ is as in Lemma \ref{c1lm} and $J > \delta/\varepsilon$.
\vskip.2cm \noindent {\bf Proof of Theorem \ref{1tm}:} If the model
has no phase transitions,  by Proposition \ref{2pn}, the set
$\mathcal{R}$ is void and $M(J,h)$ is a continuous function of $h\in
\mathbb{R}$ for each $J>0$. This yields, see (\ref{12D}) and Lemma
\ref{c3lm}, that for every $J>0$, there exists $h_*\in \mathbb{R}$,
such that $M(J,h_*)=0$. On the other hand, by (\ref{c8a}) and
(\ref{c}), (\ref{ad2}), we have in (\ref{ba3})
\[
 \varkappa \geq (\varepsilon - \delta/J) \theta (J) -
 \frac{1}{J(2 \pi)^d}
\int_{(-\pi , \pi]^d} \frac{{\rm d}p}{E(p)}.
\]
As the right-hand side  does not depend on $h$,  we pick up $J_*>0$
such that $\varkappa >0$ for all $J>J_*$. For such $J$ and $m>m_*$,
(\ref{ba}) holds. Then by Lemma \ref{abbalm}, we get a contradiction
with the supposition made at the beginning of the
proof.$\blacksquare$

We conclude this section by giving the path integral
representations\footnote{See \cite{Alb3,Alb6,KP} for more details.}
\begin{eqnarray}
\label{20} M^{\rm per}_\Lambda (J,h) & = &
\int_{\mathit{\Omega}_\Lambda} \omega_\ell (0) \mu^{\rm per}_\Lambda
({\rm d}\omega_\Lambda), \\
\mathit{\Gamma}^\Lambda_{\ell \ell'} (\tau, \tau' ) & = &
\int_{\mathit{\Omega}_\Lambda} \omega_{\ell }(\tau)\omega_{\ell'
}(\tau') \mu^{\rm per}_\Lambda ({\rm d}\omega_\Lambda).\nonumber
\end{eqnarray}
Here
\begin{equation} \label{17}
\mathit{\Omega}_\Lambda = \{\omega_\Lambda = (\omega_\ell)_{\ell \in
\Lambda} \ | \ \omega_\ell \in \mathcal{C}, \ \ {\rm for} \ {\rm
all} \ \ \ell \in \Lambda\},
\end{equation}
and $\mathcal{C}$ is the Banach space of all continuous functions
 $\omega:[0,1]\rightarrow \mathbb{R}$, such that
 $\omega(0) = \omega(1)$, equipped with the usual
supremum norm $|\cdot|_{\mathcal{C}}$. In the Hilbert space $
\mathcal{L}^2 \ \stackrel{\rm def}{=} \ L^2 ([0,1], {\rm d}\tau)$,
one defines the operator $A = - m {{\rm d}^2 }/{{\rm d }\tau^2} +
a$. Its spectrum consists of the eigenvalues
\begin{equation} \label{170}
 \lambda_k = m (2 \pi
k/\beta)^2 + a, \quad k  \in \mathbb{Z}.
\end{equation}
Thus, $A^{-1}$ is of trace class and the Fourier transform
\begin{equation} \label{160}
\int_{L^2_\beta} \exp[\imath (\phi,
\upsilon)_{\mathcal{L}^2}]\chi({\rm d}\upsilon) = \exp\left\{ -
\frac{1}{2} (A^{-1} \phi, \phi)_{\mathcal{L}^2} \right\}, \ \ \phi
\in \mathcal{L}^2
\end{equation}
defines a Gaussian measure $\chi$ on $\mathcal{L}^2$, which
obviously depends on $m$. By means of (\ref{170}), one can show that
for any $k\in \mathbb{N}$,
\begin{equation}
\label{161} \int_{\mathcal{L}^2} [\omega (\tau) -
\omega(\tau')]^{2k} \chi(d \omega) \leq \frac{2^k \Gamma
(1/2+k)}{m^k \Gamma (1/2)} \cdot |\tau - \tau'|^k_{\rm per},
\end{equation}
which by Kolmogorov's lemma, page 43 of \cite{[Sim2]}, yields that
 $\chi(\mathcal{C})=1$. Thereby, we redefine $\chi$ as a probability
measure on $\mathcal{C}$. An account of its properties may be found
in \cite{Alb3}. Thereby, the measure  in (\ref{20}) is obtained with
the help of the Feynman-Kac fromula
\begin{equation} \label{mul}
\mu^{\rm per}_{\Lambda}({\rm d}\omega_\Lambda) =  \exp\left[- I^{\rm
per}_\Lambda (\omega_\Lambda) \right]\chi_\Lambda ({\rm
d}\omega_\Lambda)/{Z^{\rm per}_\Lambda},
\end{equation}
 as a Gibbs modification of the `free measure'
\[
\chi_\Lambda ({\rm d}\omega_\Lambda) = \prod_{\ell \in \Lambda}
\chi({\rm d}\omega_\ell).
\]
Here
\begin{equation} \label{19}
I^{\rm per}_\Lambda (\omega_\Lambda)  {=} - \frac{J}{2} \sum_{\ell ,
\ell' \in \Lambda: \ |\ell - \ell'|_\Lambda=1}  (\omega_\ell ,
\omega_{\ell'})_{\mathcal{L}^2} + \sum_{\ell \in \Lambda} \int_0^1 V
(\omega_\ell (\tau)){\rm d} \tau.
\end{equation}

\section{Comments and Proof of the Lemmas}
\label{sec4}

\subsection{Comments on Propositions}

Here we discuss the proof of Propositions \ref{1pn}, \ref{4pn},
\ref{perpn}, \ref{abbalm}, and \ref{ibpn}.

 By means of
the Euclidean realization of the states (\ref{12F}), one can prove
that the sets $\{\varrho_\Lambda\}_{\Lambda \Subset \mathbb{L}}$,
$\{\varrho^{\rm per}_\Lambda\}_{\Lambda\in \mathcal{L}_{\rm box}}$,
are relatively compact in the topology which guaranties the
convergences stated in Proposition \ref{4pn}. The boundedness
mentioned in Proposition \ref{1pn} follows from the moment estimates
for Euclidean Gibbs measures proven in \cite{Alb6,KP}. The existence
of periodic Gibbs states was proven in \cite{KP}. The estimate
(\ref{a0}) follows from the FKG inequality, for the Euclidean Gibbs
measures proven in \cite{Alb3,KP}. The proof of (\ref{12L}) was
performed in \cite{KP}, see Lemma 6.4. If (\ref{ba}) holds, then the
limiting periodic Euclidean Gibbs states are nonergodic, which
certainly means a phase transition, see \cite{KP} and the references
therein. The estimate (\ref{c}) is the infrared bound, the proof of
which is standard, see \cite{BK,Kond,Alb3,KP}.

\subsection{Proof of Lemma \ref{c3lm}}

We start by proving the first line in (\ref{c4}). To this end, we
find a strictly increasing function $\phi:[h_{+}(m_0), +\infty)
\rightarrow \mathbb{R}$ such that
\begin{equation} \label{N1}
p_\Lambda^{\rm per}(J,h) \geq \phi(h) \quad {\rm for} \ \ h\geq
h_{+}(m_0).
\end{equation}
Then we use the convexity of $p^{\rm per}_\Lambda (J, \cdot)$ and
get the result in question by (\ref{12G}). Let us split the
potential $V_0$ into even and odd parts
\[
V_0(x ) = V_0^{\rm e}(x) + V_0^{\rm o}(x).
\]
Thereby, for $b>0$,  we choose $\underline{h}>0$ such that, for all
$h>\underline{h}$, $h x_\ell - V_0^{\rm e}(x_\ell)$ is an increasing
function of $x_\ell \in [-b,b]$. Set, c.f. (\ref{17}), (\ref{mul}),
\[
\mathcal{C}_b = \{ \omega \in \mathcal{C} \ | \
|\omega|_{\mathcal{C}}\leq b\}, \quad \mathit{\Omega}_\Lambda^{b} =
\{ \omega_\Lambda = (\omega_{\ell})_{\ell \in \Lambda} \ | \
\omega_\ell \in \mathcal{C}_b, \ \ \ell \in \Lambda\}, \]
\begin{equation} \label{N3}
Z^b_\Lambda (J,h) = \int_{\mathit{\Omega}^b_\Lambda}\exp\left[ -
I^{\rm per}_\Lambda (\omega_\Lambda) \right]\chi_\Lambda ({\rm
d}\omega_\Lambda).
\end{equation}
Obviously,
\begin{equation} \label{N6}
p_\Lambda^{\rm per}(J,h) \geq \frac{1}{|\Lambda|} \log Z^b_\Lambda
(J,h).
\end{equation}
 By the first GKS inequality, see e.g., Theorem 12.1 in
\cite{[Sim2]},
\begin{equation} \label{B9}
Z_{\Lambda}^{b}(J,h) \geq Z_{\Lambda}^{b}(0,h )\ \stackrel{\rm
def}{=} \  \exp\left( |\Lambda| \phi(h)\right), \quad {\rm for} \
{\rm all} \ h> \underline{h}.
\end{equation}
Here
\begin{equation} \label{N7}
\exp \left[\phi(h)\right] = \int_{\mathcal{C}_b} \exp\left\{
\int_0^1 \left[ h \omega(\tau) -   V_0( \omega(\tau))\right]{\rm
d}\tau\right\}\chi({\rm d}\omega).
\end{equation}
By Jensen's inequality, for every $\tilde{h}>0$ and $h \geq
\tilde{h}$,
\begin{equation} \label{N8}
\phi (h) \geq (h - \tilde{h}) \gamma (m,\tilde{h}) + \phi
(\tilde{h}), \quad \gamma(m,h) \ \stackrel{\rm def}{=} \ \phi'(h).
\end{equation}
By (\ref{161}), one can show, see \cite{Alb3}, that for any $m_0>0$,
the family of the corresponding measures $\{\chi\}_{|m \geq m_0}$ is
tight as measures on $\mathcal{C}$. On the other hand, the
right-hand side of (\ref{N7}) can be extended to the whole complex
plane as an entire ridge function of $h$ with the ridge being the
real axis. Thereby, for any $m_0>0$, there should exist $\tilde{h}$
such that
\[
\gamma_* (\tilde{h}) \ \stackrel{\rm def}{=} \ \inf_{m\geq m_0}
\gamma (h, \tilde{h}) >0.
\]
Then for a fixed $m_0$, we take $h_{+} (m_0) = \max \{\tilde{h},
\underline{h}\}$, which yields (\ref{N1}) and hence the first part
of (\ref{c4}). Since we have not employed any property of $V_0^{\rm
e}$, the rest of the lemma can be proven by changing the sign of $h$
and all $\omega_\ell$.

\subsection{The Main Estimate}
The proof of Lemmas \ref{c1lm}, \ref{c2lm} is based on the estimate
which we derive now.

 The path measure $\nu_h$ corresponding to the anharmonic oscillator
(\ref{2}) with the external field $h$ is defined as a probability
measure on $\mathcal{C}$ by the following expression, c.f.
(\ref{5A}), (\ref{mul}), and (\ref{19}),
\begin{equation}
\label{b8h} \nu_h ({\rm d} \omega) = \frac{1}{N_h}\exp\left[ h
\int_0^1 \omega(\tau) {\rm d}\tau - \int_0^1 V_0(\omega(\tau)){\rm
d}\tau \right]\chi({\rm d}\omega),
\end{equation}
where $1/ N_h$ is a normalization factor. For a given fixed $m_0$,
let $h_{\pm}(m_0)$ be as in Lemma \ref{c3lm}. Then for $\epsilon
>0$ and
\begin{equation} \label{BB8}
h\in [h_{-} (m_0) - \epsilon, h_{+} (m_0) +\epsilon], \quad  \ m
> m_0,
\end{equation}
by (\ref{161}) we readily get
\begin{equation} \label{162}
\int_{\mathcal{C}} [\omega (\tau) - \omega(\tau')]^{2k} \nu_h (d
\omega) \leq m^{-k} Q_k  \cdot |\tau - \tau'|^k_{\rm per}, \quad k
\in \mathbb{N},
\end{equation}
which holds, uniformly in $h$ and $m$ obeying (\ref{BB8}), with
$Q_k$ depending on $\epsilon$ only. In the sequel, we always assume
that $h$ and $m$ are chosen according to (\ref{BB8}). Since $V_0$ is
continuous and defined on the whole $\mathbb{R}$, every
finite-dimensional projection of $\nu_h$ is non-degenerate, which
yields that for every $n\in \mathbb{N}$ and $c>0$ both sets
\begin{equation} \label{b9}
C^{\pm}(n; c) \ \stackrel{\rm def}{=} \ \{ \omega \in \mathcal{C} \
| \ \pm \omega(j/n) \geq c, \ \  j = 1, \dots , n \}
\end{equation}
are such that $\nu_h[C^{\pm}(n; c)]>0$.
\begin{lemma} \label{keylm}
For every integer $n\geq2$ and any $\varepsilon >0$, there exist
$\underline{m}\geq m_0$, $c > \sqrt{\varepsilon}$, and
$B^{\pm}_\varepsilon \subset C^{\pm}(n;c)$, such that for all
$m>\underline{m}$,
\begin{equation} \label{b86}
\nu_h(B^{\pm}_\varepsilon)>0,
\end{equation}
and for all $\omega \in B^{\pm}_\varepsilon$,
\begin{equation}
\label{b85} \forall \ \tau \in [0,1]: \qquad \pm \omega (\tau) \geq
\sqrt{\varepsilon}.
\end{equation}
\end{lemma}
{\bf Proof:} Let us fix $p\in \mathbb{N}\setminus \{1\}$, $\alpha
\in (0, p-1)$, and set
\begin{equation} \label{b15}
\lambda_\vartheta (\omega) = \sup\left\{ \frac{[\omega (\tau) -
\omega(\tau')]^{2p}}{|\tau - \tau'|^\alpha_{\rm per}}\ | \ 0 <
|\tau- \tau'|_{\rm per} \leq \vartheta\right\}, \quad \vartheta \in
(0, 1).
\end{equation}
Then by the Garsia-Rodemich-Rumsey lemma, see e.g., pages 202, 203
in \cite{BY}, one has from (\ref{162})
\begin{eqnarray} \label{b16}
\int_{\mathcal{C}}\lambda_\vartheta (\omega) \nu_h(d \omega) &\leq&
\frac{2^{ \alpha + 6p + \varsigma}}{p-\alpha -1}\left(1 +
\frac{2}{\alpha}\right) m^{-p} Q_p \vartheta^{p-\alpha} \\ &
\stackrel{\rm def}{=} & m^{-p} Q_{p, \alpha} \vartheta^{p-\alpha},
\nonumber
\end{eqnarray}
where $\varsigma>0$ is an absolute constant. Now we fix $n\geq2$ and
for $c>\sqrt{\varepsilon}$, define
\begin{eqnarray*}
A(c; \varepsilon)& = & \{\omega \in \mathcal{C} \ | \ \lambda_{1/n}
(\omega) \leq (c - \sqrt{\varepsilon})^{2p}n^\alpha\},
\\ B^{\pm}_\varepsilon & = & A(c; \varepsilon)\bigcap
 C^{\pm}(n;c). \nonumber
\end{eqnarray*}
Then for any $\tau\in [0,1]$, one can pick up $j/n$, such that
\[
|\omega(\tau) - \omega(j/n)| \leq \left[ \lambda_{1/n} (\omega)
\right]^{1/2p} n ^{-\alpha /2p},
\]
which yields $\pm \omega(\tau) \geq \sqrt{\varepsilon}$ if $\omega
\in  B^{\pm}_\varepsilon$. To estimate $\nu_h(B^{\pm}_\varepsilon)$
we proceed as follows. By (\ref{b85}) and (\ref{b16}), and by the
Chebyshev inequality
\begin{eqnarray*}
\nu_h \left[\mathcal{C} \setminus A(c; \varepsilon) \right] &\leq
&\frac{1}{ (c - \sqrt{\varepsilon})^{2p} n^{\alpha}} \int
\lambda_{1/n} (\omega) {\nu}_h ({\rm d} \omega) \\ & \leq &  m^{-p}
\cdot {Q_{p,\alpha}}/{[n (c - \sqrt{\varepsilon})^2]^p}. \nonumber
\end{eqnarray*}
Now we set
\[
\sigma(n;c) = \min\left\{\nu_h\left[C^+(n;c) \right];
\nu_h\left[C^{-}(n;c) \right] \right\}.
\]
Thereby,
\begin{eqnarray} \label{b88}
\nu_h(B^{\pm}_\varepsilon) &\geq & \sigma (n;c) -
\nu_h\left[\mathcal{C} \setminus A(c; \varepsilon) \right]\\ &  \geq
& \sigma(n;c) - m^{-p} \cdot {Q_{p,\alpha}}/{[n (c -
\sqrt{\varepsilon})^2]^p}, \nonumber
\end{eqnarray}
which is positive for all
\begin{equation} \label{b23}
m > \underline{m} \ \stackrel{\rm def}{=} \ \max\left\{ m_0; \  [n
(c- \sqrt{\varepsilon})^2]^{-1}\cdot \left[ Q_{p, \alpha}/ \gamma
(n;c)\right]^{1/p}\right\}.\end{equation} $\blacksquare$ \vskip.1cm
\noindent Now we introduce
\begin{equation} \label{b10}
 \mathcal{C}\times \mathcal{C} \ni (\omega , \omega')
\mapsto Y(\omega , \omega') = \int_0^1 \omega(\tau) \omega'(\tau) d
\tau. \nonumber
\end{equation}
Then, by (\ref{b85})
\begin{equation}
\label{b12}
   \forall \omega, \omega' \in B^{\pm}_\varepsilon: \ \
 Y(\omega , \omega')
\geq \varepsilon.
\end{equation}

\subsection{The proof of the Lemmas}
{\bf Proof of Lemma \ref{c1lm}:} In the Euclidean approach, the
periodic pressure (\ref{12F}) has the following representation, see
(\ref{mul}), (\ref{19}), and (\ref{b8h}),
\begin{equation} \label{b24}
 p^{\rm per}_{\Lambda} (J,h) - p^{\rm per}_\Lambda (0,h)
 = |\Lambda|^{-1}
\log\left\{\int_{\mathit{\Omega_\Lambda}} \exp\left[{J}Y_\Lambda
(\omega_\Lambda) \right]\prod_{\ell \in \Lambda}\nu_h (d
\omega_\ell) \right\},
\end{equation}
where
\begin{equation} \label{b25}
Y_\Lambda (\omega_\Lambda) = \frac{1}{2} \sum_{\ell ,\ell' \in
\Lambda, \ |\ell - \ell'|_\Lambda = 1} Y(\omega_\ell,
\omega_{\ell'}),
\end{equation}
and $Y$ being as in (\ref{b10}). For $\pm h \geq 0$, we get from
(\ref{b24}), (\ref{b12})
\begin{eqnarray*}
   p^{\rm per}_{\Lambda} (J,h) - p^{\rm per}_\Lambda (0,h)
 & \geq & |\Lambda|^{-1}
  \log\left\{
\int_{\left(B^{\pm}_\varepsilon\right)^{|\Lambda|}} \exp\left[{J}Y
  (\omega_\Lambda) \right]\prod_{\ell
  \in \Lambda}\nu_h (d \omega_\ell) \right\}\\
& \geq & d J \varepsilon + \nu_h \left(B^{\pm}_\varepsilon  \right)
.
\end{eqnarray*}
Now we fix $\varepsilon$, $c$, $n$, and $\alpha$. Then for a given
$\delta >0$, we denote by $m_*$ the least value of $m\geq m_0$ for
which the second line in (\ref{b88}) is $\geq \exp (-\delta)$.
Thereafter, the latter estimate turns into (\ref{c2}).
$\blacksquare$ \vskip.2cm \noindent {\bf Proof of Lemma \ref{c2lm}:}
As  $p_\Lambda^{\rm per}$ is a convex function of $J$, we have
\begin{eqnarray} \label{b27}
 p^{\rm per}_\Lambda (J,h) - p_\Lambda (0,h) & = & \int_0^J
\left(\frac{\partial}{\partial t} p^{\rm per}_\Lambda (t,h)\right)
dt \\ & \leq & J \frac{\partial}{\partial J} p^{\rm per}_\Lambda
(J,h). \nonumber
\end{eqnarray}
Then by (\ref{b24}), (\ref{b25}), (\ref{b10}),
\begin{eqnarray}\label{b29}
& & \frac{\partial}{\partial J} p^{\rm per}_\Lambda (J,h) =
\frac{1}{2|\Lambda|} \sum_{\ell , \ell' \in \Lambda, \ |\ell -
\ell'|_\Lambda =1} \int_{\mathit{\Omega}_\Lambda} \omega_\ell (0)
\omega_{\ell'} (0) \mu_\Lambda^{\rm per}({\rm d}\omega_\Lambda)
\qquad
\\& & \quad  \leq \frac{1}{4|\Lambda|} \sum_{\ell , \ell' \in
\Lambda, \ |\ell - \ell'|_\Lambda =1}\int_{\mathit{\Omega}_\Lambda}
\left\{[\omega_\ell (0)]^2 + [\omega_{\ell'} (0)]^2 \right\}
\mu_\Lambda^{\rm per}({\rm d}\omega_\Lambda) \nonumber \\ & & \quad
= \frac{1}{4|\Lambda|} \sum_{\ell , \ell' \in \Lambda, \ |\ell -
\ell'|_\Lambda =1} \left\{ \varrho^{\rm per}_\Lambda \left( q_\ell^2
\right)
 + \varrho^{\rm per}_\Lambda \left( q_{\ell'}^2 \right)
\right\} = d \varrho^{\rm per}_\Lambda \left( q_\ell^2 \right),
\nonumber
\end{eqnarray}
since $\varrho^{\rm per}_\Lambda \left( q_\ell^2 \right)$ is
independent of $\ell$. Here, c.f. (\ref{b24}),
\begin{eqnarray*} \label{b30}
\mu^{\rm per}_\Lambda (d \omega_\Lambda) & = & \exp\left[ F_\Lambda
(J,h)+ JY_\Lambda (\omega_\Lambda)  \right] \prod_{\ell
\in \Lambda}\nu_h(d \omega_\ell), \\
F_\Lambda (J,h) &=& |\Lambda|\left[p_\Lambda (0,0) - p^{\rm
per}_\Lambda (J,h) \right] \nonumber
\end{eqnarray*}
 is the Euclidean local Gibbs measure which corresponds to
the Hamiltonian (\ref{9}). Thereby, we employ (\ref{b29}) in
(\ref{b27}) and get (\ref{c3}). $\blacksquare$


\end{document}